\documentclass[12pt,preprint]{aastex}
\usepackage{natbib}
\usepackage{xcolor}
\usepackage[multiple]{footmisc}
\newcommand{\eps}[1]{\mbox{log~$\epsilon$(#1)}} 
\newcommand\species[2]{#1 {\sc #2}}

\def\eg{\mbox{e.g.}}

\def\teff{\mbox{T$_{\rm eff}$}}
\def\logg{\mbox{log~{\it g}}}
\def\vmicro{\mbox{$\xi_{\rm t}$}}
\def\kmsec{\mbox{km~s$^{\rm -1}$}}
\def\onetwo{\mbox{HD 122563}}
\def\onefour{\mbox{HD 140283}}
\def\log $gf${\mbox{$\log gf$}}

\shorttitle{RR Lyr Stars}
\shortauthors{Chadid et al.}

\begin{document}

\title{THE CHEMICAL COMPOSITIONS OF VERY METAL-POOR STARS
       \onetwo\ AND \onefour; A VIEW FROM THE INFRARED}

\author{
Melike Af{\c s}ar\altaffilmark{1,2},
Christopher Sneden\altaffilmark{2},
Anna Frebel\altaffilmark{3},
Hwihyun Kim\altaffilmark{2,4},
Gregory N. Mace\altaffilmark{2},
Kyle F. Kaplan\altaffilmark{2},
Hye-In Lee\altaffilmark{5},
Hee-Young Oh\altaffilmark{4},
Jae Sok Oh\altaffilmark{4},
Soojong Pak\altaffilmark{5},
Chan Park\altaffilmark{4},
Michael D. Pavel\altaffilmark{2,6},
In-Soo Yuk\altaffilmark{4},
Daniel T. Jaffe\altaffilmark{2}
}

\altaffiltext{1}{Department of Astronomy and Space Sciences,             
                 Ege University, 35100 Bornova, {\. I}zmir, Turkey;
                 melike.afsar@ege.edu.tr}
\altaffiltext{2}{Department of Astronomy and McDonald Observatory, 
                 The University of Texas, Austin, TX 78712, USA; 
                 chris,dtj,hwihyun,mace@astro.as.utexas.edu}
\altaffiltext{3}{Kavli Institute for Astrophysics and Space Research and 
                 Department of Physics, Massachusetts Institute of Technology, 
                 Cambridge, MA 02139, USA;
                 afrebel@mit.edu}
\altaffiltext{4}{Korea Astronomy and Space Science Institute (KASI), 
                 Daejeon, Republic of Korea;
                 hyoh,ojs001,chanpark,yukis@kasi.re.kr}
\altaffiltext{5}{School of Space Research (IR Lab), Kyung Hee University
                 1732 Deokyoungdaero, Giheung-gu, Yongin-si Gyeonggi-do 
                 446-701, Republic of Korea; huynhanh7,soojong@khu.ac.kr}
\altaffiltext{6}{Institute for Astrophysical Research, Boston University, 
                 Boston, MA 02215, USA, michaeldpavel@gmail.com}

\begin{abstract}
From high resolution ($R$~$\simeq$ 45,000), high 
signal-to-noise 
($S/N$~$>$~400) spectra gathered with the Immersion Grating Infrared 
Spectrograph (IGRINS) in the H and K photometric bands, we have derived 
elemental abundances of two bright, well-known metal-poor halo stars:
the red giant \onetwo\ and the subgiant \onefour.
Since these stars have metallicities approaching [Fe/H]~=~$-$3, their
absorption features are generally very weak.
Neutral-species lines of Mg, Si, S and Ca are 
detectable, as well as those of the light odd-Z elements Na and Al.
The derived IR-based abundances agree with those obtained from 
optical-wavelength spectra.
For Mg and Si the abundances from the infrared transitions are 
improvements to those derived from shorter wavelength data.
Many useful OH and CO lines can be detected in the IGRINS \onetwo\ spectrum,
from which derived O and C abundances are consistent to those obtained 
from the traditional [\species{O}{i}] and CH features.
IGRINS high resolutions H- and K-band spectroscopy offers promising ways 
to determine more reliable abundances for additional metal-poor stars whose 
optical features are either not detectable, or too weak, or are based 
on lines with analytical difficulties.
\end{abstract}


%%%%%%%%%%%%%%%%%%%%%%%%%%%%%%%%%%%%%%%%%%%%%%%%%%%%%%%%%%%%%%%%%%%%%%%%%%
\section{INTRODUCTION\label{intro}}
%%%%%%%%%%%%%%%%%%%%%%%%%%%%%%%%%%%%%%%%%%%%%%%%%%%%%%%%%%%%%%%%%%%%%%%%%%

Element production in the early Galaxy commenced quickly with the 
explosive deaths of high-mass stars.
Their ejecta were gathered up to form the low-mass very metal-poor
stars that we observe today.
These stars are powerful tracers of the onset of Galactic nucleosynthesis 
because they live long enough to still be observable today.
Their atmospheres preserve information on the chemical history of the very
early Milky Way.
The spectroscopic task is to derive accurate abundances for as many
elements as possible of important nucleosynthetic groups (LiBeCNO,
$\alpha$, light odd-Z, Fe-peak, and neutron-capture) in low metallicity 
stars to constrain our ideas about early Galactic chemical evolution.

Low-metallicity stars exhibit weak and uncrowded absorption spectra, 
which makes abundance analyses relatively easy.  
However, this asset is also a liability, because the number of lines 
detectable for many elements drops precipitously as metallicities 
decline from solar to the ultra-metal-poor regime.
It is difficult to find any absorption features at all in
the record-setting halo stars with [Fe/H]\footnote{
We adopt the standard spectroscopic notation \citep{wal59} that for
elements A and B,
[A/B] $\equiv$ log$_{\rm 10}$(N$_{\rm A}$/N$_{\rm B}$)$_{\star}$ $-$
log$_{\rm 10}$(N$_{\rm A}$/N$_{\rm B}$)$_{\odot}$.
We use the definition
\eps{A} $\equiv$ log$_{\rm 10}$(N$_{\rm A}$/N$_{\rm H}$) + 12.0, and
equate metallicity with the stellar [Fe/H] value.}~$<$~$-$7, 
such as SMSS~J031300.36-670839.3 (\eg, Figure~1 of \citealt{kel14}).
Indeed, in the abundance literature for stars with [Fe/H]~$\lesssim$~$-$2.5, 
many elements are represented only by their resonance or very low-excitation 
lines.
Species relying on 1-3 lines include for example \species{Na}{i} 
(the D-lines); \species{Al}{i} (3961~\AA); and \species{Mn}{i} 
(the 4030~\AA\ resonance triplet).
This is unfortunate, because abundances derived from very strong 
low-excitation lines sometimes are very discordant with those derived from
weaker high-excitation lines.

High-resolution spectra in different wavelength regimes can provide 
additional absorption features to improve the reliability of many abundances
in metal-poor stars.
The vacuum ultraviolet contains ionized species to complement many of 
the elements that have only neutral-species features in the optical 
spectral region.  
But only a small number of low metallicity stars have ever been observed
at high spectral resolution in the $UV$ with the Hubble Space Telescope.
That list of stars is unlikely to grow significantly in the near future.

The infrared wavelength regime contains many neutral-species transitions of the more 
abundant light and Fe-group elements.
High-resolution $IR$ spectra now can be obtained with ground
based facilities, such as the VLT/CRIRES\footnote{
CRyogenic high-resolution InfraRed Echelle Spectrograph; \\
http://www.eso.org/sci/facilities/paranal/instruments/crires.html;
currently undergoing an upgrade.}
\citep{kae04}, and the Gemini/GNIRS\footnote{
Gemini Near Infra-Red Spectrograph;
http://www.gemini.edu/sciops/instruments/gnirs.}
\citep{eli06}.
An important $H$-band high-resolution ($R$~$\simeq$ 22,500) 
survey of $>$10$^5$ stars in the APOGEE \citep{maj10}\footnote{
Apache Point Observatory Galactic Evolution Experiment;
https://www.sdss3.org/surveys/apogee.php.}
experiment is currently underway, with a southern hemisphere extension planned
for the near future.
The new high-resolution H and K band Immersion Grating 
Infrared Spectrograph (hereafter IGRINS; \citealt{yuk10})\footnote{
http://www.as.utexas.edu/astronomy/research/people/jaffe/igrins.html.}
was commissioned in May 2014 at the McDonald Observatory for use with the 
2.7m Smith Telescope.
This instrument has been employed for many studies, but early use has
concentrated on young stars, the interstellar medium, and various types
of emission nebulae.

Very metal-poor stars have not been prime targets for $IR$ spectroscopy
due to their extreme line weakness.
But as part of the commissioning effort for IGRINS, we obtained 
spectra for two well-studied halo stars with [Fe/H]~$<$~$-$2: 
the red giant \onetwo\ and the subgiant \onefour.
In this paper, we consider what abundance information can be extracted 
from these high-resolution, very high signal-to-noise ($S/N$) spectra that
encompass the complete photometric H and K bands.
In \S2 we describe IGRINS, the observations of our stars, and basic data
reductions.
\S3 brings in previously collected high-resolution optical data for \onetwo\
and \onefour\ in order to derive new model atmospheric parameters for these
stars.
The derivation of abundances from the IGRINS spectra are contained in \S4,
and discussion of these results appears in \S5.

%%%%%%%%%%%%%%%%%%%%%%%%%%%%%%%%%%%%%%%%%%%%%%%%%%%%%%%%%%%%%%%%%%%%%%%%%%
\section{OBSERVATIONS AND REDUCTIONS\label{observations}}
%%%%%%%%%%%%%%%%%%%%%%%%%%%%%%%%%%%%%%%%%%%%%%%%%%%%%%%%%%%%%%%%%%%%%%%%%%

We obtained high-resolution, high signal-to-noise ($S/N$~$>$~400, per resolution element) spectra 
of \onetwo\ and \onefour\ with IGRINS during an instrument commissioning
run on the night of May 23, 2014. 
In Table~\ref{tab-obs}, we give basic data and observational parameters
for these stars.

Extended descriptions of IGRINS are given in \cite{yuk10} and
\cite{par14}; here we highlight some of its characteristics.
Usage of a silicon immersion echelle grating enables this spectrograph 
to achieve high resolving power, 
$R$~$\equiv$ $\lambda/\Delta\lambda$~$\simeq$ 45,000
in both H and K wavelength regions. 
IGRINS employs two 2K$\times$2K pixel Teledyne Scientific and Imaging 
H2RG detectors and it has continuous coverage of both H and K bands 
1.48$-$2.48~$\mu$m, 14,800$-$24,800~\AA) together in a 
single exposure, which allows observers to obtain all the data simultaneously. 
There is only a small gap between H and K bands of about 100~\AA.
For point sources such as stars, standard $IR$ observational techniques are 
employed during data acquisition. 
That is, four individual exposures of targets and telluric standard stars 
are taken with an ABBA sequence. 
A and B nod positions are placed along the 1$\times$15 arcsecond slit 
with a separation of 7~arcseconds. 
The application of this technique allows one to properly subtract the 
sky and telescope backgrounds, yielding a clean signal of the targets.
Flat field, ThAr lamp and sky exposures are also collected during 
each night as needed.

For point source observations IGRINS achieves $S/N$~$\sim$~100
per resolution element in a one hour exposure on a K~=~10.5 mag target.
Our stars are much brighter (Table~\ref{tab-obs}), and with a total exposure 
of 960 seconds we achieved $S/N$~$>$~400 from photon statistics alone
throughout the H and K bands.

To transform the raw data frames into final echelle multi-order (23 in the
H-band, 21 in the K-band) spectra
we processed the data with the IGRINS reduction pipeline package
PLP2\footnote{The pipeline package PLP2 has been developed 
by the pipeline team led by Jae-Joon Lee at Korea Astronomy and 
Space Science Institute and by Soojong Pak at Kyung Hee University.}\footnote{
Currently available at: https://github.com/igrins/plp/tree/v2.0}.
This suite of tasks in PLP2 automatically performs a variety of reduction 
procedures: flat fielding, background removal, order extraction, distortion 
correction, and wavelength calibration.
Sky (OH) emission lines were cancelled by subtracting the AB pairs. 
Once the sky contributions were removed, Th-Ar lamp exposures were used 
for the initial wavelength calibration. 
Then the calibration was refined by fitting the OH emission lines in
the OH lines appearing in the original ``uncancelled'' stellar spectra

Telluric water vapor is the main contaminant in both H and K bands, 
and CO$_2$ becomes more effective especially in the latter.  
Precise removal of the telluric features can be a challenge and needs 
special attention. 
Therefore we also observed telluric standards (fast-rotating hot stars, 
from mid-B to late A-type) at air masses that were very close to those
of our target stars.
We used the IRAF\footnote{
IRAF is distributed by the National Optical Astronomy Observatory, which
is operated by the Association of Universities for Research in Astronomy
(AURA) under cooperative agreement with the National Science Foundation.} 
task $telluric$ to divide out the telluric features. 
This task was performed interactively for each echelle order in the H- 
and K-band spectra.

\begin{figure}
\epsscale{0.75}
\plotone{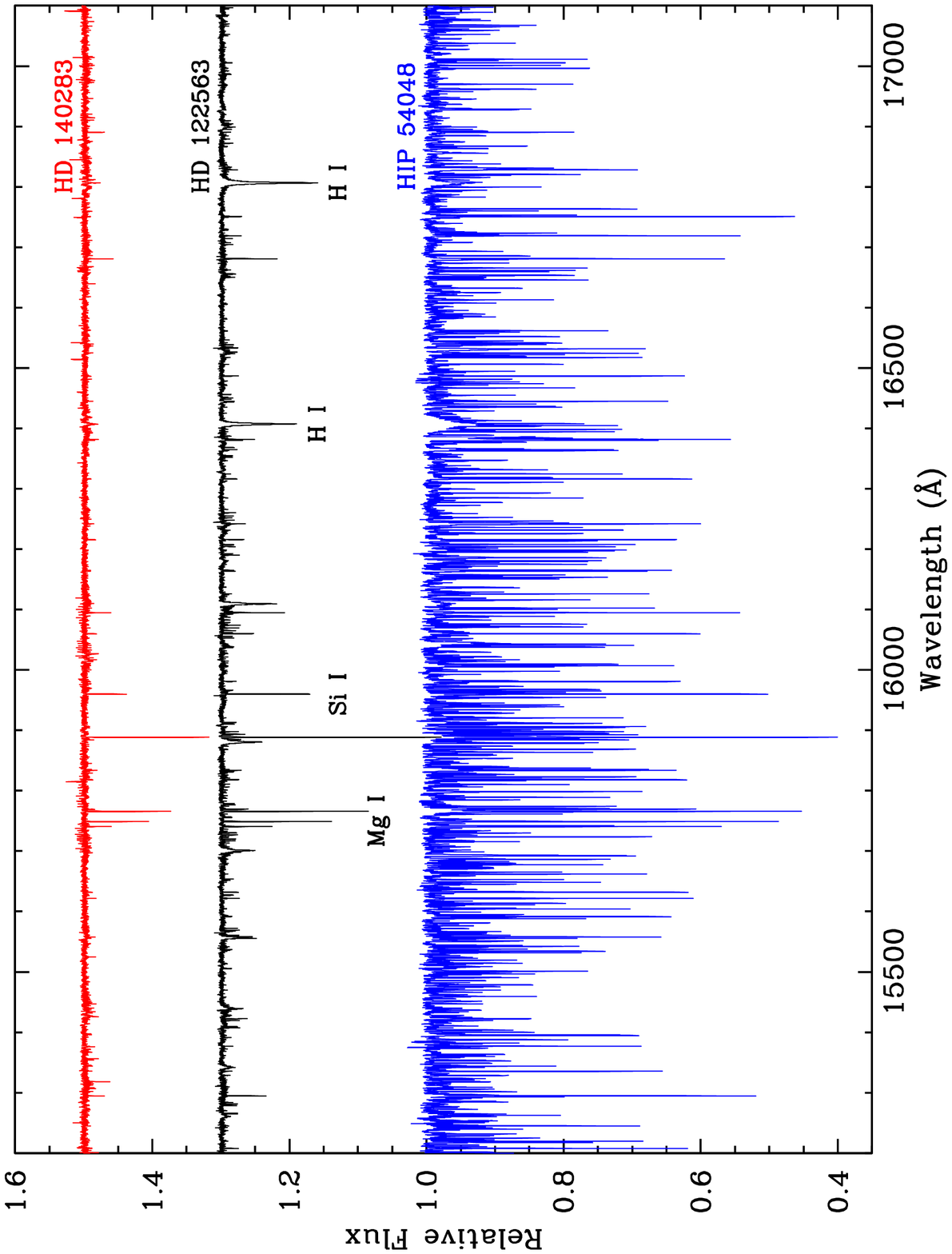}
\caption{
\label{hspec}
\footnotesize
IGRINS H-band spectra of the very metal-poor subgiant \onefour\ (red).  
The very metal-poor giant \onetwo\ (black), and
the metal-rich red horizontal-branch star HIP~54048 (blue). 
The relative flux scale is correct for HIP~54048, and for plotting clarity we
have shifted the spectra of \onetwo\ and \onefour\ vertically with 
additive constants.
We have trimmed the spectra at the low and high wavelength edges to avoid
spectrum regions at the H-band edges with severe telluric contamination.
A few prominent features are labeled in the plot, but all of the lines 
more than a few percent deep in the target star spectra are true
stellar absorptions.
}
\end{figure}

\begin{figure}
\epsscale{0.75}
\plotone{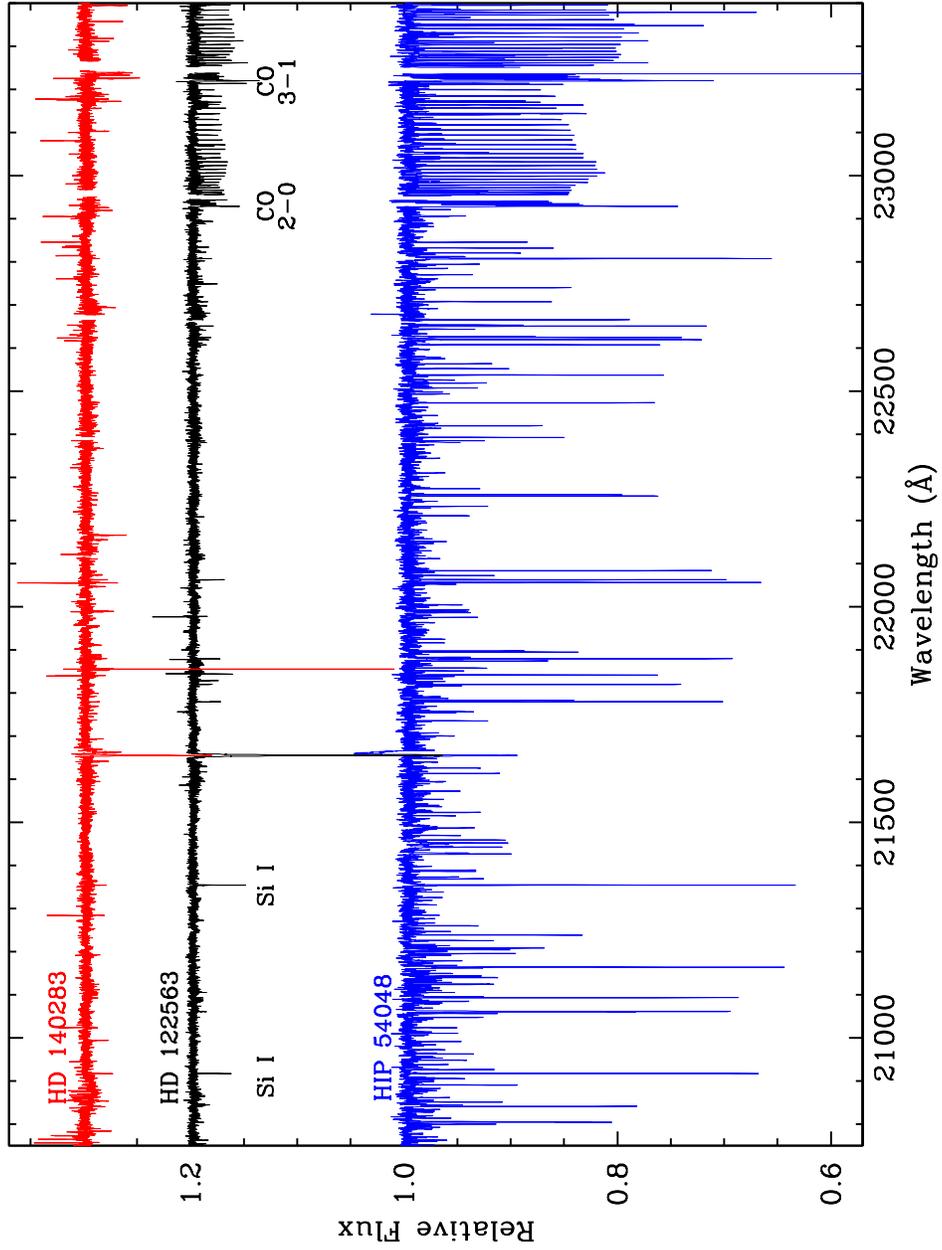}
\caption{
\label{kspec}
\footnotesize
IGRINS K-band spectra of 
\onefour\ (red).
\onetwo\ (black), and
HIP~54048 (blue). 
Lines and labels are as in Figure~\ref{hspec}.
}
\end{figure}

The wavelength-calibrated, continuum-normalized, telluric-line-excised 
echelle spectra were then merged into a single continuous spectrum with 
the IRAF task $scombine$.
The final H- and K-band spectra ready for analysis are shown in 
Figures~\ref{hspec} and \ref{kspec}.
We adopt wavelength units of \AA ngstroms in these and subsequent figures 
because we are combining results from optical and $IR$ spectra.
These figures are restricted in wavelength to those parts of the H and K
bands for which good telluric line removal was achieved; the heavily
sky-contaminated (and useless for analysis) band edges are not shown.
To demonstrate the very weak-lined nature of \onetwo\ and \onefour\ we
also plot the reduced IGRINS spectrum of the red horizontal-branch
star HIP~54048.
This is a solar-metallicity Galactic thin disk red horizontal-branch
star with atmospheric parameters (\teff, \logg, [Fe/H], \vmicro) = 
(5100~K, 2.65, 0.00, 1.2~\kmsec), which were derived by \cite{afs12} from 
high-resolution optical spectroscopy.
The HIP~54048 spectrum is crowded, with many overlapping strong spectral
features.
In contrast, \onetwo\ and especially \onefour\ have few absorption features
with central depths as large as 10\%.
It is clear that both high spectroscopic $R$ and $S/N$ are needed to
effectively employ H- and K-band spectroscopy on very metal-poor stars.

%%%%%%%%%%%%%%%%%%%%%%%%%%%%%%%%%%%%%%%%%%%%%%%%%%%%%%%%%%%%%%%%%%%%%%%%%%
\section{MODEL ATMOSPHERES AND OPTICAL-REGION ABUNDANCES\label{models}}
%%%%%%%%%%%%%%%%%%%%%%%%%%%%%%%%%%%%%%%%%%%%%%%%%%%%%%%%%%%%%%%%%%%%%%%%%%

Our program stars have been the targets of high-resolution spectroscopy
for more than half a century.
\onetwo\ was the first very metal-poor red giant to have an extensive
abundance analysis \citep{wal63}, and this honor for very metal-poor
subgiants went to \onefour\ \citep{cha51}.
These stars have more than 25 abundance studies each since the year
2000 that are cited in the SAGA Database \citep{sud08}\footnote{
http://saga.sci.hokudai.ac.jp/wiki/doku.php.}.
However, in order to make an internally self-consistent comparison 
of abundances newly derived from $IR$ spectra with those from 
optical-wavelength spectra, we began by re-determining model parameters
for both stars.

The optical spectrum for \onetwo\ was obtained with the McDonald Observatory
2.7m Smith Telescope and echelle spectrograph \citep{tul95} configured to
yield $R$~$\simeq$~60,000, wavelength coverage 
3500~\AA~$\leq$ $\lambda$~$\leq$ 9000~\AA, and $S/N$~$\sim$~125 at
$\lambda$~$\sim$~4500~\AA, increasing to $\sim$250 at 
$\lambda$~$\sim$~6500~\AA.
This spectrum was originally obtained by \cite{wes00} to provide a 
contrasting star to the neutron-capture-rich low metallicity giant
HD~115444.
We measured equivalent widths ($EW$) for lines of \species{Fe}{i}, 
\species{Fe}{ii}, 
\species{Ti}{i}, \species{Ti}{ii},
and neutral-species
lines of light elements Na, Mg, Al, Si, and Ca, using the spectroscopic 
manipulation code $SPECTRE$ \citep{fit87}.
Lines chosen for analysis were those with wavelengths $\lambda$~$>$~4500~\AA\
(to avoid significant line blending) that were detectable, isolated, and 
have well-tested transition probabilities: 
\species{Ti}{i}, \cite{law13}; \species{Ti}{ii}, \cite{woo13};
\species{Fe}{i} \cite{den14}, \cite{ruf14}, \cite{obr91};
\species{Fe}{ii} and the remaining species, the NIST\footnote{
NIST is the National Institute of Standards and Technology (NIST) Atomic 
Spectra Database; http://physics.nist.gov/PhysRefData/ASD/} 
atomic line database \citep{kra14}.
The Fe and Ti line parameters and $EW$s are listed in Table~\ref{tab-feti}
and those of other species are in Table~\ref{tab-others}.

Since \onefour\ is 1150~K hotter and has a gravity more than two dex 
higher than \onetwo, it has a much weaker-lined spectrum.  
Therefore we were able to measure $EW$s of only about half as many atomic 
lines for this star.
Our main spectrum for this task was originally part of an investigation of
the blue metal-poor star CS~29497-030 \citep{iva05}.
It was obtained with the Keck~I High Resolution Echelle Spectrometer 
(HIRESb, \citealt{vog94}). It has $R$~$\simeq$~40,000, wavelength coverage
3050~\AA~$\leq$ $\lambda$~$\leq$ 5900~\AA, and $S/N$~$\sim$~250 at
$\lambda$~$\sim$~4500~\AA, increasing to $\sim$300 at
$\lambda$~$\sim$~5800~\AA.  
To measure a few lines redward of 5900~\AA\ we also used a spectrum gathered
as part of the large-sample very-metal-poor halo star survey of
\cite{roe14}.
This spectrum, obtained with the Las Campanas Observatory Magellan Inamori 
Kyocera Echelle (MIKE) spectrograph \citep{ber03}, has 
$R$~$\simeq$~35,000,  5000~\AA~$\leq$ $\lambda$~$\leq$ 8500~\AA, 
and $S/N$~$\sim$~140 at $\lambda$~$\sim$~6500~\AA.
The $EW$s measured for these spectra are listed in Tables~\ref{tab-feti} 
and \ref{tab-others}.

We derived atmospheric parameters for \onetwo\ and \onefour\ by determining
line-by-line abundances from the $EW$s and trial model atmospheres 
interpolated from the ATLAS model atmosphere $\alpha$-enhanced
grid \citep{kur11}\footnote{
http://kurucz.harvard.edu/grids.html; model interpolation software was
developed by Andy McWilliamd and Inese Ivans.}.
The abundance computations were carried out with the version of the
line analysis code MOOG \citep{sne73} that includes scattering in the
continuum opacity and source function \citep{sob11}.
Model parameters were adjusted by using standard criteria that line
abundances of \species{Fe}{i} show no trend with excitation potential
$\chi$ and equivalent width $EW$ (for \teff\ and \vmicro), that the mean
abundances of neutral and ionized species agree for Fe and Ti (for \logg),
and that the assumed model metallicity are consistent with the derived [Fe/H].

After model iterations, we obtained atmospheric parameters 
effective temperature, surface gravity, microturbulent velocity, and model
metallicity (\teff/\logg/\vmicro/[M/H]) = \\
(4500~K/0.80/2.20~\kmsec/$-$2.90) for \onetwo\
and (5650~K/3.40/1.70~\kmsec/$-$2.70) for \onefour.
These parameters are consistent with past analyses of these two stars.
Straight averages of all the SAGA entries for each star yield 
parameters (4570~K, 1.1, $-$2.7) for \onetwo\
and (5730~K, 3.6, $-$2.5) for \onefour.
Our derived metallicities are about 0.2~dex lower than these means of previous
studies.
A small part ($\sim$0.05~dex) of this metallicity offset is 
attributable to the use of the scattering-included version of MOOG.
The more important reason is that our derived temperatures are lower by 
about 75~K than the average of entries in the SAGA database.
Those previous studies that derive \teff\ values similar to ours typically
also derive [Fe/H] similar to ours.

We then derived abundances of light elements Na, Mg, Al, Si, Ca from all
of their neutral-species transitions that have log~$gf$ values given in
the NIST database and $EW$s greater than 2.5~m\AA.
For the purposes of this paper, we ignored all other elements beyond
Ca (Z~=~20) because they are not detected in our IGRINS spectra.
Because there is no single literature source for the transition probabilities
of these elements that encompasses the optical and $IR$ spectral domains,
we opted to adopt the log~$gf$ values from NIST, ignoring for our work
the NIST quality estimates of these values.
Derived abundances from individual transitions are given in 
Table~\ref{tab-others}, and species means are given in Table~\ref{tab-means}.

%%%%%%%%%%%%%%%%%%%%%%%%%%%%%%%%%%%%%%%%%%%%%%%%%%%%%%%%%%%%%%%%%%%%%%%%%%
\section{THE INFRARED ANALYSIS\label{infrared}}
%%%%%%%%%%%%%%%%%%%%%%%%%%%%%%%%%%%%%%%%%%%%%%%%%%%%%%%%%%%%%%%%%%%%%%%%%%

Using our high quality IGRINS H- and K-band spectra, 
we were able to determine abundances of light elements C, O, Na, Mg,
Al, Si, S and Ca in \onetwo, and Mg, Al and Si in \onefour.
All of the abundance computations were accomplished with synthetic/observed
spectrum matches.
As mentioned in \S\ref{models}, the much weaker-lined spectrum of 
\onefour\ yielded only about half as many abundances compared to \onetwo.
In this section we discuss the atomic/molecular data choices and 
describe some details of the abundance analysis.

\subsection{Atomic and Molecular Data\label{linedata}}

To create synthetic spectrum line lists we combined the best atomic and 
molecular literature data from several sources. 
We generated the line list by initially adopting atomic and CO molecular 
lines in selected regions of the H- and K-bands from the \cite{kur11} 
database\footnote{
Available at http://kurucz.harvard.edu/linelists.html}. 
Then we added CN and OH molecular lines from recent lab publications
by \cite{sne14} and \cite{bro15}.

Bandheads of the CO ground electronic state first 
overtone ($\Delta$$v$ = 2) system are prominent in K-band spectra 
of cool stars (Figure~\ref{kspec}), and second overtone lines 
($\Delta$$v$ = 3) occur throughout their H-band spectra.
We initially adopted CO transition data from the Kurucz database,
which were generated from the calculations of \citealt{goo94}.
We then tested these line lists by calculating synthetic CO spectra 
of the solar photosphere with these lines, adopting a solar 
model atmosphere interpolated in the \cite{cas03} grid with 
parameters (\teff/\logg/\vmicro/[M/H]) = \mbox{(5780~K/4.44/0.85~\kmsec/0.00)}
and well-determined C and O abundances 
from optical features (\eg, \citealt{asp09} and references therein).
These synthetic spectra were compared to the solar photospheric flux 
spectrum of \cite{wal03}.
The H-band $\Delta$$v$ = 3 ``second overtone'' lines showed satisfactory 
synthetic/observed spectrum agreement with the abundances deduced 
from the optical region. 
However the K-band $\Delta$$v$ = 2 computed lines were too strong compared
to the observed ones, and an increase in C abundance of $\sim$0.2~dex 
was needed to resolve the discrepancy.

We also synthesized the spectrum of Arcturus \citep{hin05}\footnote{
Available at ftp://ftp.noao.edu/catalogs/arcturusatlas/}, 
adopting an interpolated ATLAS
model atmosphere with parameters \mbox{(4285~K/1.66/1.74~\kmsec/$-$0.52};
\citealt{ram11}) and the C and O abundances of \cite{sne14}.
Very similar results were obtained: good synthetic/observed agreement for
the second overtone bands but the same underprediction of the first 
overtone bands.
Resolution of this discrepancy is beyond the scope of this work, but
it has been explored in detail in other investigations (\eg, \citealt{ayr13}
and references therein), with the probable culprit being inadequacies in
1-dimensional atmospheric modeling.
For the present purposes, we simply decided to increase the log $gf$ 
values of the $\Delta$$v$ = 2 first overtone lines by 0.15 dex. 
We then re-applied these new log $gf$ values to the solar and Arcturus
spectra; resulting C abundances yielded good agreement between
the optical and H-band abundances.

To find suitable atomic lines we began with the lists of
\cite{hin95} that accompany their infrared Arcturus atlas, and then searched
for them in our \onetwo\ and \onefour\ spectra.
Transition probabilities for the chosen lines were collected from the NIST 
database, when available. 
The NIST database does not include log~$gf$ recommendations for $IR$ 
\species{Ca}{i} lines, so we applied reverse solar analysis to determine 
their log $gf$ values. 
For this task we again used the \cite{wal03} high-resolution infrared 
solar flux spectrum.
\species{Ca}{i} transition probabilities determined in this manner were 
then tested on the \cite{hin05} spectrum of Arcturus.
Good agreement was found with the Arcturus Ca abundance derived by
\cite{ram11}.

\subsection{Abundance Analysis\label{abanalysis}}

The near-infrared spectra of cool stars contain many molecular
lines, especially OH and CO.
Even at low metallicities the possibility exists that atomic lines 
may be blended. 
Thereby, we decided to apply the spectrum synthesis technique to all 
available infrared lines (Table~\ref{tab-others}) to derive the elemental 
abundances of C, O, Na, Mg, Al, Si, S and Ca. 

Transitions of $\alpha$-elements Mg, Si, S, and Ca
were detected in our IGRINS spectrum of \onetwo, and Mg and Si were 
represented by lines in both H- and K-bands in both program stars; 
see Table~\ref{tab-others}. 
These $IR$ lines significantly improve the abundances deduced from the
optical transitions.
Derived mean abundances are presented in Table~\ref{tab-means}.

In Figure~\ref{si1spec}, we display 10 \species{Si}{i} lines in the
\onetwo\ spectrum.
The top panels show the violet 3905 and 4102~\AA\ features, the only 
strong \species{Si}{i} in very metal-poor stellar spectra.
The 4102~\AA\ line's continuum is the wing of H$\beta$, and the 3905~\AA\ line
is partially blended.
Moreover, a standard abundance analysis of the 3905~\AA\ line yields 
\teff-dependent Si abundances (\eg, \citealt{sne08b} and references therein).
All other optical-region \species{Si}{i} transitions are very weak in
\onetwo, as illustrated in the four middle panels of Figure~\ref{si1spec}.
Only the 5684~\AA\ line has a depth of more than a few percent.
These lines are rarely encountered in studies of very metal-poor stars.
In contrast, we easily detect more than a dozen \species{Si}{i} lines
with reasonably well-determined transition probabilities as judged
in the NIST compilation.
The optical-region Si detection problems for \onefour\ are even more severe, 
as only three lines (3905, 4102, and 5684~\AA) lines could be measured in 
our spectra.
But our $IR$ abundance is based on nine lines.
Additionally, the line-to-line scatters for both stars are significantly
smaller in the $IR$ lines than in their optical counterparts.
The abundances of Si derived from our $IR$ spectra are 
clearly more reliable than those from the optical data.

Many of the same comments on \species{Si}{i} could be applied to the 
\species{Mg}{i} lines in our program stars.
Strong $IR$ lines can be easily used for Mg abundances, as we illustrate
in the top panel of Figure~\ref{synobs}.
Here we also show synthetic spectrum matches to all the features in this
wavelength interval.
Inspection of these \species{Mg}{i} line profiles suggests that $EW$ 
analysis could have accomplished the abundance derivation task.
But other $IR$ spectral features would require synthetic spectrum calculations,
and so we have applied the synthetic spectra to all $IR$ transitions.
For \species{Mg}{i} the optical transitions are more competitive
with the $IR$ ones in quantity and quality, with the line-to-line abundance
scatters being comparable (Table~\ref{tab-means}.
The increase in Mg abundance reliability comes from the
merging of results from both optical and $IR$ bands.

Three lines of \species{S}{i} can be detected in \onetwo.
However, as seen in the bottom two panels of Figure~\ref{synobs}, they 
are extremely weak, with central depths of $\lesssim$1\%.
Unsurprisingly, these lines proved to be too weak for detection in \onefour.
From our synthetic spectrum calculations we derived an abundance
for HD 122563 of $<$\eps S$>$~=~4.72) from the 15470, 15478 and 22707~\AA\ 
\species{S}{i} lines, or [S/H]~=~$-$2.40, [S/Fe]~=$+$0.52.  
\onetwo\ was included in the S abundance survey of \cite{spi11}, who 
studied two stronger \species{S}{i} lines at 9228 and 9237~\AA.
They derived [S/Fe]~=~$+$0.42 with a 3D+NLTE calculation, and a purely
LTE abundance roughly 0.3~dex larger.
To our knowledge, NLTE computations have not been done for our lines,
but since they have high excitation energies, $\chi$~$\gtrsim$~8.0~eV,
we suspect that they might not be negligible.
Future investigation of this issue will be worthwhile.
However, our S/Fe overabundance for \onetwo\ is in good agreement with
previous studies of low metallicity stars (\citealt{nis04}, 
\citealt{spi11}, \citealt{caf14}).
Nearly all published S abundances have been based on the \species{S}{i} 
lines shortward of 10,000~\AA.
IGRINS now gives us a new set of lines to conduct large-sample surveys of
this element.

Six $IR$ \species{Ca}{i} lines were detectable in our \onetwo\ 
(Table~\ref{tab-others})
spectrum, but all were too weak to be seen in \onefour.
There is good agreement between the \onetwo\ Ca abundances derived from the 
optical and infrared regions (Table~\ref{tab-means}), and indicate 
[Ca/Fe]~$\simeq$~0.35, consistent with Ca overabundances observed
in most low-metallicity stars.

Our IGRINS spectra contain a few transitions of odd-Z species \species{Na}{i}
and \species{Al}{i}.
Two weak Na lines could be detected for \onetwo\ in the K-band 
(Table~\ref{tab-others}). 
The mean abundance, $\langle$[Na/Fe]$\rangle$~=~$-$0.11 is 0.2~dex smaller
than the mean from the three optical-wavelength lines (Table~\ref{tab-means}),
but the line-to-line abundance scatters are large, $\sigma$~$\geq$0.16.
The overall mean Na abundance is compatible with previously reported 
values for very metal-poor stars (\eg, \citealt{cay04}, \citealt{yon13}, 
and references therein).

In our optical-wavelength spectra only the \species{Al}{i} resonance
doublet at 3944 and 3961~\AA\ can be detected, and the 3944~\AA\ line
has long been known \citep{mag83} to be severally contaminated by CH.
The 3961~\AA\ line indicates a large Al deficiency in our stars, 
[Al/Fe]~$\lesssim$~$-$0.9 in our LTE analysis.
Although we could not detect other well-studied optical-region 
higher-excitation \species{Al}{i} doublets (\eg, 6696/6698~\AA, 7835/7836~\AA, 
or 8772/8773~\AA), abundances reported in the literature on metal-poor stars 
from these lines computed in LTE analyses are [Al/Fe]~$\sim$~0.
This clash between the resonance lines and other \species{Al}{i} lines
is well-documented, \eg, \cite{fra86}, \cite{rya96}, and is a clear
example of NLTE effects in resonance line formation in metal-poor stars.
\cite{bau97} performed a detailed NLTE analysis of Al in subgiants and
main sequence stars over a range of metallicities, showing that NLTE Al
abundance corrections are much larger for the resonance lines than
that higher-excitation ones.
For \cite{cay04}'s survey of very metal-poor giants, which included \onetwo,
a uniform NLTE correction of $+$0.65 was adopted for the 3961~\AA\ line.
If we apply that to our LTE value of [Al/Fe]~=~$-$0.70 (Table~\ref{tab-means})
we obtain $-$0.05, in agreement with [Al/Fe]~=~$-$0.04 from the four
$IR$ lines.
Assuming that a similar correction applies to the LTE abundance derived
from the 3961~\AA\ line in \onefour\ yields similar agreement with those 
derived from the two $IR$ lines, [Al/Fe]~$\simeq$~$-$0.3.
  
Many individual OH molecular transitions can be found in the IGRINS 
H-band spectrum of \onetwo, and many CO transitions in its K-band spectrum.
The persistence of detectable OH in this very low metallicity giant is
due to the combined effects of: \textit{(a)} the well-documented O 
overabundance (many studies beginning with \citealt{lam74}, [O/Fe]~=~$-$0.6); 
\textit{(b)} the weak formation of the competing double-metal molecule CO; 
and \textit{(c)} the (relative) strengthening of hydride molecules with 
decreasing metallicity (due to increasing gas pressure).
Using the new OH ro-vibrational line lists of \cite{bro15}, we
derived O abundances from 51 OH lines (Table~\ref{tab-others}),
which yielded a mean abundance [O/Fe]~=~$+$0.92 ($\sigma$~=~0.05;
Table~\ref{tab-means}).
This value is somewhat larger than the abundance that we derived from the two 
optical-spectrum [\species{O}{i}] lines, [O/Fe]~=~$+$0.78.
However, recently \cite{dob15} have carried out detailed 3D atmospheric
modeling of [\species{O}{i}] and OH ro-vibrational features in red giants
including \onetwo,
finding little change of the forbidden lines compared to 1D analyses
but decreases of 0.2$-$0.3~dex in derived O abundances from the OH lines.
Detailed comparisons with their work are beyond the scope of this work,
but a correction of this order of magnitude would bring the derived O
abundance into satisfactory agreement with O abundance
estimates for \onetwo.
Due to the large temperature sensitivity of molecular features, we were
unable to detect either OH or CO in \onefour.

As described in previous section, we raised the log $gf$ values of K-band CO 
lines by 0.15~dex in order to have a self-consistent C abundance from 
both optical (CH) and infrared (CO) transitions. 
For CO we adopted the O abundance derived from OH, and varied the C 
abundance in the syntheses. The C abundance could only be determined 
for HD 122563 using the $\Delta$$v$ = 2 first overtone lines that appear in the K-band. Second 
overtone line ($\Delta$$v$ = 3) strengths are too weak to be detected in either stars.
We derive a larger abundance from CO by 0.13~dex than from CH.
However, CO molecular line strengths have larger temperature sensitivity 
than do the CH or OH lines, because of the large dissociation difference 
between the two molecules.
The molecular Saha equation includes the dissociation energy in the
exponential term exp($-$D$_{0}$/kT), or in logarithmic units 
$-$D$_0\theta$, where $\theta$~$\equiv$~5040/T. 
The dissociation energies of these molecules of D$_0$(CH)~=~3.47~eV,
D$_0$(OH)~=~4.39~eV, and D$_0$(CO)~=~11.09~eV.
For \onetwo\ a 100~K \teff\ change from our assumed \teff\ would alter
the derived O abundance from OH by $\simeq$0.11~dex,
alter the C abundance from CH by $\simeq$0.09~dex, but will change the 
derived C abundance from CO by $\simeq$0.28~dex.
A small temperature decrease would bring agreement to the CH- and CO-based C 
abundances.
We therefore suggest that to first approximation both C abundance indictors
are in accord.
The same molecular Saha argument will easily show why neither OH or CO
infrared bands can be detected in \onefour, whose \teff\ is 1150~K hotter.

Finally, we have considered the effects on our derived 
abundances for \onetwo\ that would occur if we adopt the model atmospheric
parameters derived by \cite{cre12}.  
Those authors derived the angular diameter for \onetwo\ from interferometry,
and combined that value with photometric and astrometric data to
recommend \teff\ = 4598~K, \logg\ = 1.60, and [Fe/H] = $-$2.6.
We re-derived abundances with this model, finding a metallicity similar
to theirs and abundance ratios [X/Fe] from atomic lines very little 
changed from our values.
The abundances of C estimated from CH and O from OH molecular features
were also very little affected by the model change.  
The higher \teff\ of the \cite{cre12} model yields weaker CH and OH lines, 
while the much larger \logg\ strengthens these lines; the two effects
counter each other almost completely.
However, O abundances derived from [\species{O}{i}] lines becomes about a factor
of two larger than that derived from OH due to the large gravity sensitivity
of [\species{O}{i}].
Additionally, CO lines have very large sensitivity to \teff\ (see the
discussion in \ref{abanalysis}), so adoption of the \cite{cre12} model
would yield C abundances again about a factor of two larger than we have 
derived.
We note that with our lower cooler, lower gravity model we get roughly
the same C and O abundances from all relevant species, but that
is only a consistency argument.
Further study of \onetwo\ models would be welcome.

%%%%%%%%%%%%%%%%%%%%%%%%%%%%%%%%%%%%%%%%%%%%%%%%%%%%%%%%%%%%%%%%%%%%%%%%%%
\section{DISCUSSION AND CONCLUSIONS\label{results}}
%%%%%%%%%%%%%%%%%%%%%%%%%%%%%%%%%%%%%%%%%%%%%%%%%%%%%%%%%%%%%%%%%%%%%%%%%%

We used the new McDonald Observatory IGRINS high-resolution
spectrograph to investigate the $IR$ spectra of two very metal-poor 
halo stars, \onetwo\ and \onefour. 
Very few neutral-species transitions are detectable, mainly $\alpha$-elements.
Although our targets are well studied in the optical region,
we decided to re-derive the model atmosphere parameters in order to have
internal self-consistency in optical and $IR$ analyses.
We used these model atmospheres to determine the abundances of species 
from lines that have been identified in the $IR$ spectrum of Arcturus.
We were able to determine the abundances of C, O, Na, Mg, Al, Si, S and 
Ca in \onetwo, while only Mg, Al and Si were detectable in \onefour\ 
due to its higher surface temperature compared to \onetwo.

\begin{figure}
\epsscale{0.8}
\plotone{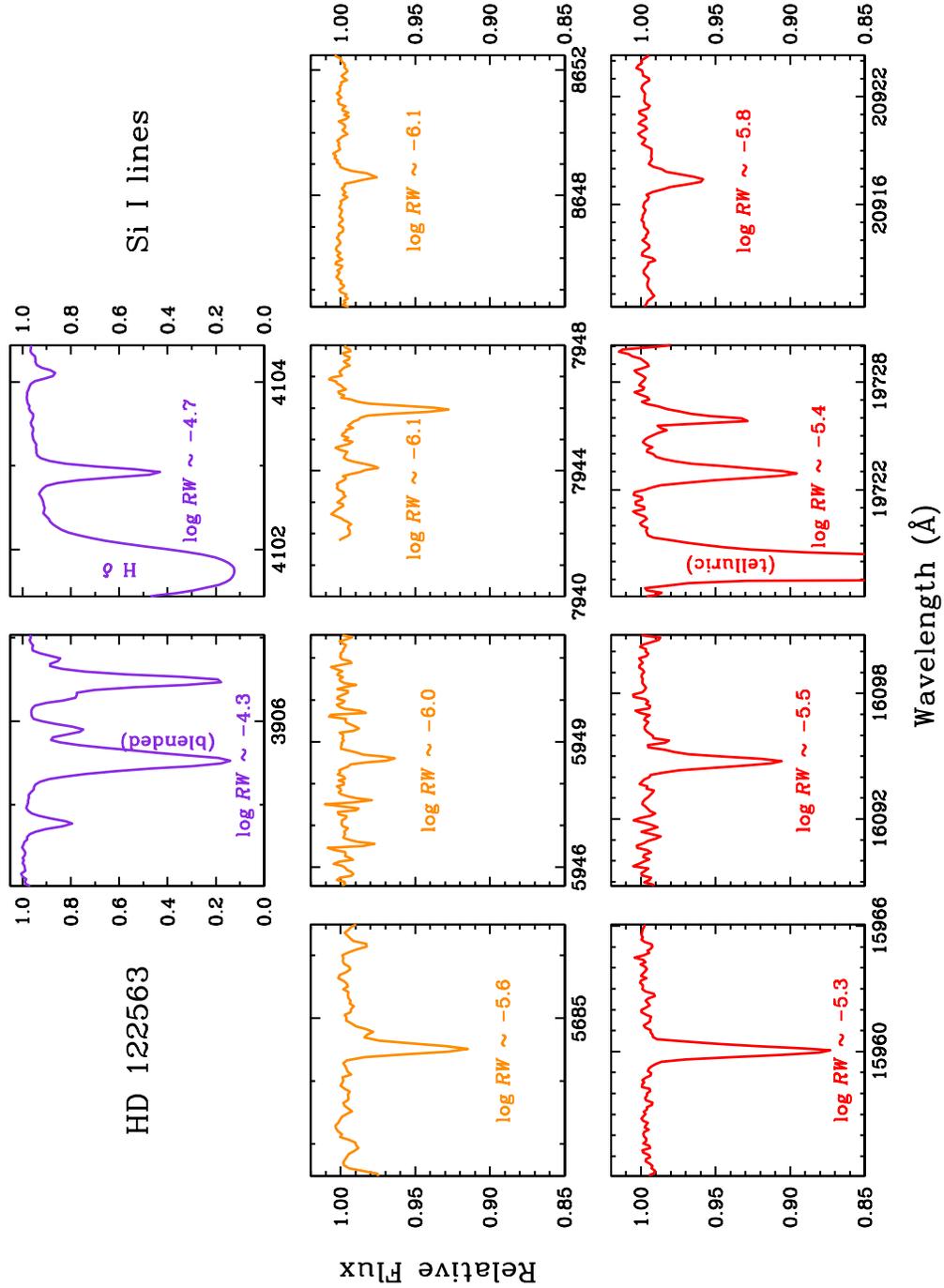}
\caption{
\label{si1spec}
\footnotesize
Small spectral regions surrounding 10 \species{Si}{i} lines in \onetwo.
In each panel we have recorded the approximate value of the \species{Si}{i} 
line reduced width in logarithmic units, log~$RW$~$\equiv$~log($EW/\lambda$),
which indicates the line strength.
Vertical lines are drawn at the line wavelengths.
The top panels display in violet the two strong lines usually used for
Si abundance studies in very low metallicity stars.
The middle panels show in orange four other typical 
optical-region lines, and the bottom panels show in red four representative 
H- and K-band lines.}
\end{figure}

\begin{figure}
\epsscale{0.9}
\plotone{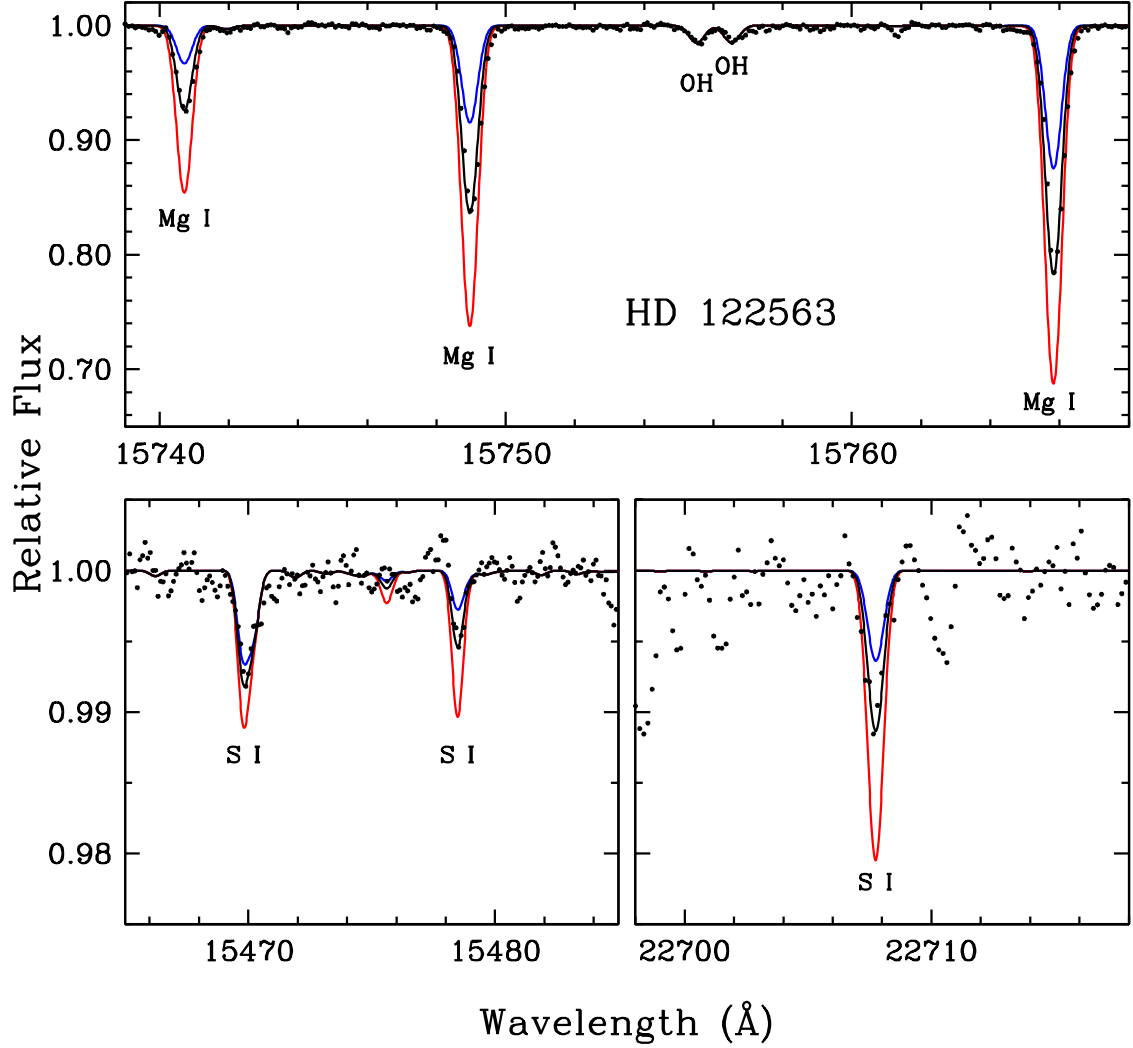}
\caption{
\label{synobs}
\footnotesize
Observed and synthetic spectra of \species{Mg} (top panel) and 
\species{S}{i} (bottom panels) lines in the H- and K-bands.
In each panel the small filled circles show the observed spectrum, and
the black line represents the synthetic spectrum with an abundance that 
best matches the transitions in that panel.
In the top panel the red and blue lines correspond to synthetic spectrum 
abundance choices that are offset from the best fit by $\pm$0.4~dex.
In the bottom panels the red and blue lines are for synthetic spectrum 
abundances offset by $\pm$0.3~dex.
}
\end{figure}

The abundances of $\alpha$-elements determined from the 
$IR$ spectra were generally in good agreement with their optical counterparts.
Many of the Mg and Si absorption lines in the IGRINS spectra of both \onetwo\
and \onefour\ are strong (Figure~\ref{si1spec}), and yielded more reliable 
abundances than that of obtained from the visible spectra. 

Sulfur is one of the $\alpha$-elements that is produced 
via oxygen burning during the core-collapse of massive stars.
It plays an important role in deciphering the Galactic alpha-element 
chemical evolution. 
The expected flat behavior observed in the [$\alpha$/Fe]-[Fe/H] plots is 
still a matter of debate, especially due to limited blend-free optical 
lines used to investigate this relation. On the other hand,
we could detect three extremely weak \species{S}{i} lines in the \onetwo\ 
H- and K-band spectra: 15470, 15478 and 22707~\AA.  
The mean abundance derived from these lines indicates that sulphur is enhanced 
about 0.5 dex as expected for an $\alpha$-element in very-metal poor stars 
(\eg, \citealt{kac15}). 
To our knowledge, these H- and K-band \species{S}{i} lines have never 
been studied for our targets. 
In fact, the 22707~\AA\ sulphur line has been used for the first time here
to determine the sulphur abundance of such a very-metal poor star. 
With IGRINS now, we have gained more useful 
sulphur lines to investigate its complex behaviour throughout the Galaxy.

We were also able to detect two odd-Z elements, Na and Al.
Both elements were only available in the \onetwo\ spectrum. 
The abundance of Na could be determined, with a significant line-to-line
scatter, for \onetwo\ from two \species{Na}{i} lines identified in 
the K-band (Table~\ref{tab-others}). 
However, Na did not reveal itself in the spectrum of \onefour. 
The abundances of aluminum determined for both stars yielded a solar 
value, which is compatible with the NLTE-corrected abundances obtained 
from the resonance lines at 3944 and 3962 ~\AA\ (e.g. \citealt{and08})
for very metal-poor stars. 
This promising result indicates that 
$IR$ \species{Al}{i} lines might be used for NLTE-free abundance 
determination, especially for very metal-poor stars that suffer from severely
contaminated aluminum lines in the blue part of the optical region. 
More definitive conclusions await further investigation of the 
$IR$ \species{Al}{i} lines in a large sample of very metal-poor stars.

Looking to future high-resolution $IR$ spectroscopic studies of very 
metal-poor stars, what could be the low metallicity limit for useful 
information?
Here we consider only red-giant stars like \onetwo\ due to their
stronger-lined spectra than subgiants or main-sequence stars at 
similar metallicities. 
One of the stronger lines in our atomic list is \species{Si}{i} at 
15888~\AA. 
Its central depth is over 30\% in the \onetwo\ spectrum. 
When we assume a S/N~$\simeq$~100 and adopt an approximate value of 
[Si/Fe]~$\simeq$~$+$0.50 for very metal-poor stars with atmospheric parameters 
similar to those \onetwo,  the 15888~\AA\ line appears to be still detectable 
with a central depth of about 8\% for an extremely metal-poor star 
([Fe/H]~$\simeq$~$-$4). 
This simple experiment shows us the capacity of IGRINS to detect at
least a few species in red giants even more metal-poor than \onetwo.

OH and CO lines are very weak in \onetwo, with $\sim$1$-$4\% central depths.
However, there are many that can be detected in its spectrum. 
We used 51 OH and 38 CO lines to form our abundance means for O and C.
Therefore, as a detection experiment we selected one K-band echelle order
and first shifted 16 CO lines that appear in it from their rest positions 
to a common scale referenced to their line centers. 
We overplot these shifted absorption features in panel (a) of
Figure~\ref{stack3}.
We then co-added them to produce an ``average'' CO line; this
is displayed in the panel (b) of Figure~\ref{stack3}.
A similar line averaging exercise was done for $UV$ CO 
lines by \cite{fre07}.
While the original spectrum has high signal-to-noise, $S/N$~$\simeq$~450
in this wavelength region, the co-added spectrum has $S/N$~$\simeq$~1000.
The average line, although just 3\% deep, can be easily analyzed.

We have computed synthetic spectra with parameters 
characteristic of the 16 CO lines under consideration (excitation energy 
$\chi$~=0.11, transition probability $-$5.72), and plot the results 
in panel (c) of Figure~\ref{stack3}.
For this test the absolute abundance of CO was not important.
We matched the observed average line as closely as possible, and
then repeated the synthesis several times with abundances shifted
in abundance steps of 0.3~dex.
Visual inspection of the observed/synthetic spectrum matches,and in the 
observed $minus$ computed spectrum differences shown in panel (d)
of the figure, suggest that the double-metal molecule CO can be detected 
in red-giant stars with metallicities at least 0.3~dex lower than \onetwo.

Detections of atomic transitions will not be easy at lower 
metallicities than \onetwo.
As mentioned above, the strongest line in our H- and 
K-band spectra of this star is \species{Si}{i} 15888~\AA.
Our numerical experiments suggested that this line might be detectable in 
stars with metallicities as low as [Fe/H]~$\sim$~$-$4. 
The atmospheric parameters of \onetwo\ place it close to the very 
metal-poor red-giant tip.
The most luminous red giant members of the lowest metallicity globular 
clusters have temperatures and gravities as small as \teff~$\sim$~4250~K 
and \logg~$\sim$~0.0.
Such stars would have stronger-lined spectra than \onetwo\ at 
[Fe/H]~$\sim$~$-$3, thus would be attractive IGRINS targets. 
However, stars bright enough for IGRINS observations are not plentiful.
It may be as interesting to explore the spectra of somewhat higher metallicity
stars; our search for detectable transitions in \onetwo\ and \onefour\
suggests that other species such as \species{K}{i} may exhibit useful
lines for stars with [Fe/H]~$\sim$~$-$2.

In this paper, we have shown that many light elements have 
neutral transitions that can be usefully applied to abundance analyses of 
very metal-poor stars.
These spectral features are consistent with and in several cases appear
to be more reliable abundance indicators than transitions in the optical
spectral range.
Future IGRINS observations of stars spanning the metal-poor metallicity
domain should be undertaken.

\acknowledgments

This study has been supported in part by NSF grant AST-1211585 to C.S.,
by The Scientific and Technological Research Council of Turkey 
(T\"{U}B\.{I}TAK, project No. 112T929) to M.A, and by NSF-CAREER
grant AST-1255160 tp A.F.
This work used the Immersion Grating Infrared Spectrograph 
(IGRINS) that was developed under a collaboration between the University 
of Texas at Austin and the Korea Astronomy and Space Science Institute 
(KASI) with the financial support of the US National Science Foundation 
under grant AST-1229522, of the University of Texas at Austin, and of the 
Korean GMT Project of KASI.

%%%%%%%%%%%%%%%%%%%%%%%%%%%%%%%%%%%%%%%%%%%%%%%%%%%%%%%%%%%
\clearpage
\bibliography{chrisbib}
%%%%%%%%%%%%%%%%%%%%%%%%%%%%%%%%%%%%%%%%%%%%%%%%%%%%%%%%%%%

\clearpage

\clearpage

\clearpage

\clearpage

\clearpage
\begin{figure}
\epsscale{0.9}
\plotone{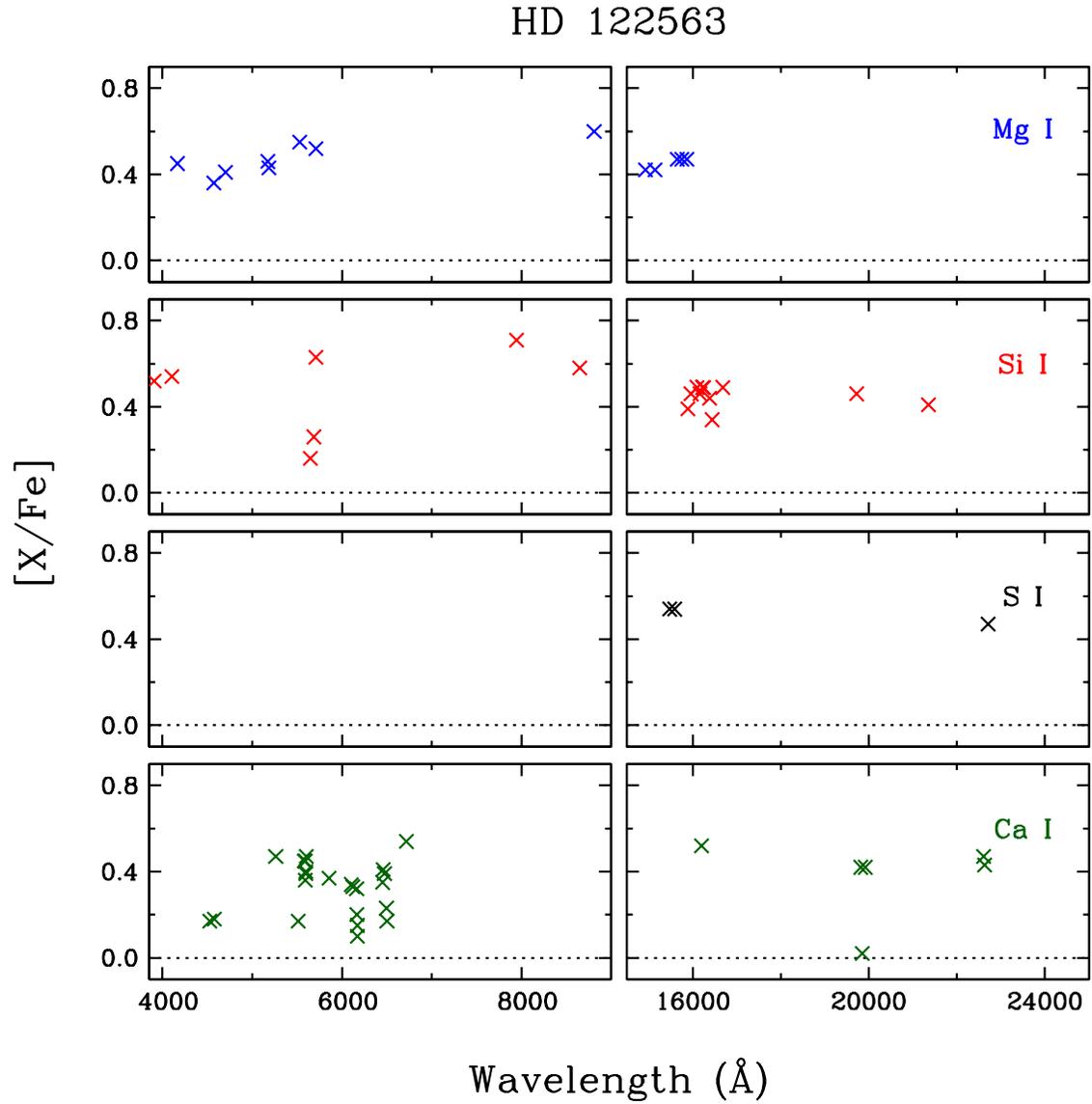}
\caption{
\label{alphas}
\footnotesize
Relative abundance ratios [X/Fe] of $\alpha$ elements (Mg, Si, S, and
Ca from top to bottom panels) in \onetwo.
The left-hand  panels are for abundances derived from the optical-region 
spectrum, and the right-hand panels are for IGRINS-based abundances.
}
\end{figure}

\clearpage
\begin{figure}
\epsscale{0.9}
\plotone{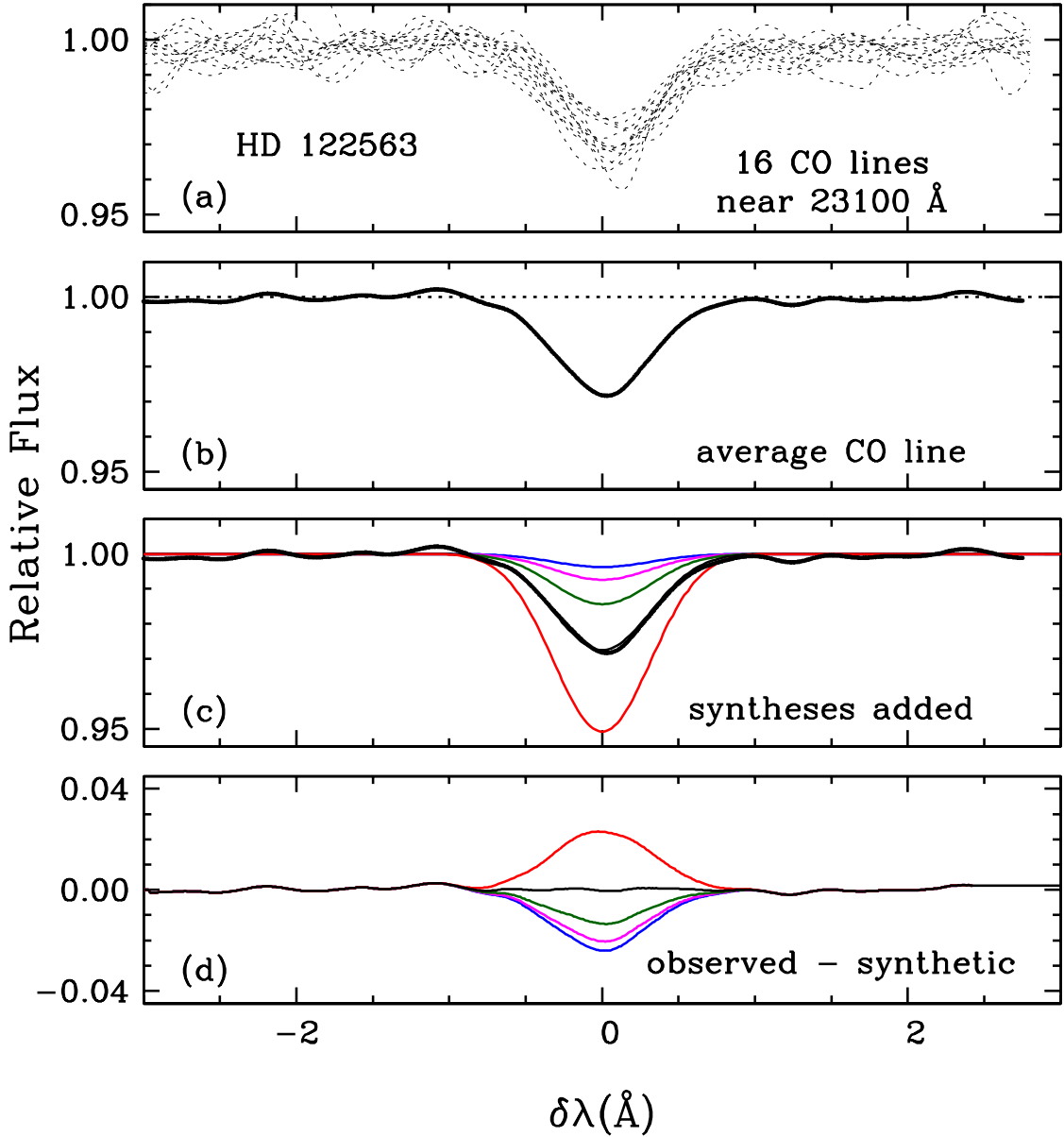}
\caption{
\label{stack3}
\footnotesize
Observed and synthetic spectra of CO first-overtone lines 
in \onetwo.  Panel (a): 16 individual lines near 23100~\AA\ are shown, shifted 
together in wavelength space by subtraction of their central wavelengths.
Panel (b): the co-addition of these 16 lines is shown.
Panel (c): the co-added observed spectrum is again plotted,
along with five synthetic spectra generated with the model atmosphere
adopted for \onetwo\ and a single CO line with a mean excitation potential
and transition probability of the co-added observed lines.
The synthesis drawn with a black line, nearly indistinguishable from 
the observed line, has C and O abundances chosen to best match the
observation.
The other synthetic spectrum lines are:  red for a C 
abundance 0.3~dex larger than the best fit, green for 0.3 smaller, 
magenta for 0.6 smaller, and blue for 0.9 smaller.
Panel (d): the difference be observed and computed spectra.
}
\end{figure}

\clearpage
\begin{center}
\begin{deluxetable}{lccccccc}
\tabletypesize{\footnotesize}
\tablewidth{0pt}
\tablecaption{Program Star Parameters and Observations\label{tab-obs}}
\tablecolumns{8}
\tablehead{
\colhead{Star}                & 
\colhead{RA\tablenotemark{a}} & 
\colhead{Dec}                 &  
\colhead{$V$}                 &  
\colhead{$H$}                 &  
\colhead{$K$}                 &  
\colhead{Exposure}            & 
\colhead{$S/N$}               \\  
\colhead{}                    &     
\colhead{(2000)}              &     
\colhead{(2000)}              &   
\colhead{}                    &   
\colhead{}                    &    
\colhead{}                    &
\colhead{(sec)}               &
\colhead{}                    
}
\startdata
HD 122563 & 14 02 31.85	& $+$09 41 09.95 & 6.19	& 3.76 & 3.73 & 960 & $>$400 \\
HD 140283 & 15 43 03.10 & $-$10 56 00.60 & 7.21	& 5.70 & 5.59 & 960 & $>$400 \\
\enddata

\tablenotetext{a}{Positions and magnitudes taken from the SIMBAD database,
http://simbad.u-strasbg.fr/simbad/}

\end{deluxetable}
\end{center}

\begin{center}
\begin{deluxetable}{lcrrrrrr}
\tabletypesize{\footnotesize}
\tablewidth{0pt}
\tablecaption{Parameters of Fe and Ti Lines\label{tab-feti}}
\tablecolumns{8}
\tablehead{
\colhead{$\lambda_{air}$}               &
\colhead{Species}                       &
\colhead{$\chi$}                        &
\colhead{log $gf$}                      &
\colhead{$EW_{122563}$}                 &
\colhead{$EW_{140283}$}                 &
\colhead{[X/Fe]\tablenotemark{a}}       &
\colhead{[X/Fe]\tablenotemark{a}}       \\
\colhead{(\AA)}                         &
\colhead{}                              &
\colhead{(eV)}                          &
\colhead{}                              &
\colhead{(m\AA)}                        &
\colhead{(m\AA)}                        &
\colhead{}                              &
\colhead{}                     
}
\startdata
4449.14 & Ti {\sc i} &  1.890 &    0.47 &  6.4 & \nodata &    0.14 & \nodata \\
4450.89 & Ti {\sc i} &  1.880 &    0.32 &  3.6 & \nodata &    0.02 & \nodata \\
4453.31 & Ti {\sc i} &  1.430 & $-$0.03 &  9.3 & \nodata &    0.26 & \nodata \\
4455.32 & Ti {\sc i} &  1.440 &    0.13 &  8.7 & \nodata &    0.08 & \nodata \\
4457.43 & Ti {\sc i} &  1.460 &    0.26 & 15.5 & \nodata &    0.25 & \nodata \\
4512.73 & Ti {\sc i} &  0.830 & $-$0.40 & 14.0 &   2.0   &    0.09 &  0.19   \\
4518.02 & Ti {\sc i} &  0.820 & $-$0.25 & 19.0 &   3.3   &    0.08 &  0.26   \\
4527.30 & Ti {\sc i} &  0.810 & $-$0.45 & 14.4 &   2.7   &    0.12 &  0.35   \\
4533.24 & Ti {\sc i} &  0.850 &    0.54 & 51.9 &  15.5   & $-$0.05 &  0.22   \\
4534.78 & Ti {\sc i} &  0.830 &    0.35 & 42.5 &  11.1   & $-$0.03 &  0.23   \\
\enddata

\tablenotetext{a}{The solar abundances are taken from \cite{asp09};
for Fe I and Fe II, the abundances are expressed as [Fe/H].}

\vspace*{0.1in}
(This table is available in its entirety in a machine-readable form in the 
online \\ journal.  
A portion is shown here for guidance regarding its form and content.)

\end{deluxetable}

\end{center}

\begin{center}
\begin{deluxetable}{lcrrrrrr}
\tabletypesize{\footnotesize}
\tablewidth{0pt}
\tablecaption{Parameters of Other Species Lines\label{tab-others}}
\tablecolumns{8}
\tablehead{
\colhead{$\lambda_{air}$}               &
\colhead{Species}                       &
\colhead{$\chi$}                        &
\colhead{log $gf$}                      &
\colhead{$EW_{122563}$}                 &
\colhead{$EW_{140283}$}                 &
\colhead{[X/Fe]\tablenotemark{a}}       &
\colhead{[X/Fe]\tablenotemark{a}}       \\
\colhead{(\AA)}                         &
\colhead{}                              &
\colhead{(eV)}                          &
\colhead{}                              &
\colhead{(m\AA)}                        &
\colhead{(m\AA)}                        &
\colhead{}                              &
\colhead{}                     
}
\startdata
 6300.31 & [O {\sc i}]  &  0.000 & $-$9.72 &  syn   & \nodata &    0.76 & \nodata \\
 6363.78 & [O {\sc i}]  &  0.020 & $-$0.19 &  syn   & \nodata &    0.79 & \nodata \\
 5889.95 &  Na {\sc i}  &  0.000 &    0.11 &  182.0 & 120.0   &    0.27 &    0.22 \\
 5895.92 &  Na {\sc i}  &  0.000 & $-$0.19 &  155.0 &  88.0   &    0.15 & $-$0.01 \\
 8183.26 &  Na {\sc i}  &  2.100 &    0.24 &   15.0 & \nodata & $-$0.08 & \nodata \\
 8194.79 &  Na {\sc i}  &  2.100 &    0.54 &   30.5 & \nodata &    0.00 & \nodata \\
22056.43 &  Na {\sc i}  &  3.191 &    0.29 &   syn  & \nodata & $-$0.23 & \nodata \\
23379.14 &  Na {\sc i}  &  3.753 &    0.54 &   syn  & \nodata &    0.02 & \nodata \\
 4167.17 &  Mg {\sc i}  &  4.350 &   -0.75 &   50.8 &  25.8   &    0.45 &    0.35 \\
 4571.10 &  Mg {\sc i}  &  0.000 & $-$5.62 &   82.5 &   7.0   &    0.36 &    0.21 \\
\enddata

\tablenotetext{a}{The solar abundances are taken from \cite{asp09}.}

\vspace*{0.1in}
(This table is available in its entirety in a machine-readable form in the 
online \\ journal. 
A portion is shown here for guidance regarding its form and content.)

\end{deluxetable}

\end{center}

\begin{center}
\begin{deluxetable}{lrrrrrr}
\tabletypesize{\footnotesize}
\tablewidth{0pt}
\tablecaption{Mean Abundances\label{tab-means}}
\tablecolumns{7}
\tablehead{
\colhead{Species}                       &
\colhead{[X/Fe]}                        &
\colhead{$\sigma$}                      &
\colhead{num}                           &
\colhead{[X/Fe]}                        &
\colhead{$\sigma$}                      &
\colhead{num}                           \\
\colhead{}                              &
\colhead{122563}                        &
\colhead{122563}                        &
\colhead{122563}                        &
\colhead{140283}                        &
\colhead{140283}                        &
\colhead{140283}                        
}
\startdata
CH-vis      
            & $-$0.28 &    0.03 &       \scriptsize{CH-band} & \nodata & \nodata & \nodata \\
CO-IR       
            & $-$0.15 &    0.06 &      38 & \nodata & \nodata & \nodata \\ 
$[$\species{O}{i}$]$-vis   
            &    0.78 &    0.02 &       2 & \nodata & \nodata & \nodata \\ 
OH-IR       
            &    0.92 &    0.06 &      51 & \nodata & \nodata & \nodata \\ 
\species{Na}{i}-vis    
            &    0.09 &    0.16 &       4 &    0.11 &    0.16 &       2 \\
\species{Na}{i}-IR     
            & $-$0.11 &    0.18 &       2 & \nodata & \nodata & \nodata \\ 
\species{Mg}{i}-VIS    
            &    0.47 &    0.08 &       8 &    0.26 &    0.06 &       7 \\
\species{Mg}{i}-IR     
            &    0.45 &    0.03 &       5 &    0.27 &    0.11 &       6 \\
\species{Al}{i}-vis    
            & $-$0.70 & \nodata &       1 & $-$0.99 & \nodata &       1 \\
\species{Al}{i}-IR     
            & $-$0.04 &    0.06	&       4 & $-$0.29 &    0.00 &       2 \\
\species{Si}{i}-vis    
            &    0.49 &    0.20 &       7 &    0.51 &    0.20 &       3 \\
\species{Si}{i}-IR     
            &    0.45 &    0.05 &      11 &    0.33 &    0.05 &       9 \\
\species{S}{i}-vis     
            & \nodata & \nodata & \nodata & \nodata & \nodata & \nodata \\ 
\species{S}{i}-IR      
            &    0.52 &    0.04 &       3 & \nodata & \nodata & \nodata \\ 
\species{Ca}{i}-vis    
            &    0.32 &    0.13 &      23 &    0.36 &    0.09 &      14 \\
\species{Ca}{i}-IR     
            &    0.38 &    0.18 &       6 & \nodata & \nodata & \nodata \\ 
\species{Ti}{i}-vis    
            &    0.12 &    0.09 &      38 &    0.27 &    0.07 &      13 \\
\species{Ti}{ii}-vis   
            &    0.29 &    0.07 &      39 &    0.34 &    0.08 &      19 \\
\species{Fe}{i}-vis    
            & $-$2.92\tablenotemark{a}
            &    0.12 &     143 & $-$2.71 &    0.05 &      66 \\
\species{Fe}{ii}-vis   
            & $-$2.98\tablenotemark{a}
            &    0.10 &      12 & $-$2.77 &    0.06 &       9 \\
\enddata

\tablenotetext{a}{For Fe only the value is [Fe/H].}

\end{deluxetable}

\end{center}
 
\end{document}